\documentclass[aps, reprint,pra, longbibliography]{revtex4-2}
\usepackage{dcolumn}
\usepackage[utf8]{inputenc}
\usepackage{amsmath}
\usepackage{amsfonts}
\usepackage{amssymb}
\usepackage{graphicx}
\usepackage{hyperref}
\usepackage{xcolor}

\newcommand{\ket}[1]{\ensuremath{\left|{#1}\right\rangle}}
\newcommand{\bra}[1]{\ensuremath{\left\langle{#1}\right |}}
\newcommand{\braket}[2]{\ensuremath{\left\langle{#1}|{#2}\right\rangle}}
\newcommand{\proj}[1]{\ensuremath{\left|{#1}\right\rangle\left\langle{#1}\right |}}
\newcommand{\avg}[1]{\ensuremath{\left\langle{#1}\right\rangle}}

\newcommand{\mb}{\mathbf}
\newcommand{\bs}{\boldsymbol}

\begin{document}

\title{Apparent quantum paradoxes as simple interference: Quantum violation of the pigeonhole principle and exchange of properties between quantum particles}

\author{Raul Corr\^ea} \author{Pablo L. Saldanha}\email{saldanha@fisica.ufmg.br}
\affiliation{Departamento de F\'isica, Universidade Federal de Minas Gerais, Belo Horizonte, MG 31270-901, Brazil}

 %\altaffiliation[Also at ]{Physics Department, XYZ University.}%Lines break automatically or can be forced with 

\date{\today}% It is always \today, today,

\begin{abstract}
It was recently argued that the pigeonhole principle, which states that if three pigeons are put into two pigeonholes then at least one pigeonhole 
must contain more than one pigeon, is violated in quantum systems [Y. Aharonov \textit{et al.}, PNAS \textbf{113}, 532 (2016)]. An experimental verification of this effect was recently reported [M.-C. Chen \textit{et al.}, PNAS \textbf{116}, 1549 (2019)]. In another recent experimental work, it was argued that two entities were observed to exchange properties without meeting each other [Z.-H. Liu \textit{et al.}, Nat. Commun. \textbf{11}, 3006 (2020)]. Here we describe all these proposals and experiments 
as simple quantum interference effects, where no such dramatic conclusions appear. Besides demystifying some of the conclusions of the cited works, we also present physical insights for some interesting behaviors present in these treatments. For instance, we associate the anomalous particles 
behaviors in the quantum pigeonhole effect to a quantum interference of force.
\end{abstract}

%\pacs{}

\maketitle

\section{Introduction}

Many quantum paradoxes were proposed with the concept of weak values \cite{aharonov88,dressel14} recently, attributing strange properties to a system between pre- and postselection. In the quantum Cheshire cat effect \cite{aharonov13,denkmayr14}, for instance, it is stated that a photon 
could be separated from its polarization or a neutron from its spin. Investigations of the past of a photon in an interferometer \cite{vaidman13,danan13} concluded that the photon had passed through places where it could not have passed. In the quantum pigeonhole effect \cite{aharonov16,waegell17,chen19,reznik20} it is argued that it is possible to put three quantum particles in two boxes, without two particles being in the same box. Finally, it has been recently argued that two photons can exchange their polarization without meeting each other \cite{das20,liu20}, in an analogy 
with an exchange of grins between quantum Cheshire cats. 

One thing that is not always clearly stated in many of the quantum paradoxes cited above is that their striking conclusions depend on a particular 
realistic interpretation of quantum mechanics. If we adopt a more pragmatic, non-realistic view, no paradox arises in the proposed gedanken and experimental configurations. In this sense, some of the above paradoxes have been explained as simple interference effects, with no mention of paradoxical behaviors, including the quantum Cheshire effect \cite{correa15,atherton15} and the past of a quantum particle in an interferometer 
\cite{saldanha14,bartkiewicz15,englert17}.

More generally, the interpretation of the weak values themselves has also 
been subject of debate. As mentioned, the proponents of the weak value concept advocate for their realistic interpretation, in which the value accessed by means of a pointer weakly interacting with the system of interest (weak 
measurement) reveals some underlying property of the system \cite{aharonov88,aharonov13}. It is under this realistic interpretation of the weak values that the mentioned quantum paradoxes appear \cite{aharonov13,denkmayr14,vaidman13,danan13,aharonov16,waegell17,chen19,reznik20,das20,liu20}. In this view, even the extraction of a weak value without the use of a pointer, by counting detections, should reveal the underlying 
property of the system \cite{denkmayr14,chen19,liu20}. On the other side of the spectrum, there are those that discuss the  weak values as being connected to a perturbation on the pointer state \cite{leggett89, sokolovski15, 
sokolovski18}, in which the anomalous weak values (those lying outside the eigenvalues range of the associated observable) appear as a feature of quantum mechanics allowed by the complex probability amplitude between state transitions and interference effects. There is also the investigation 
of which statistical information the weak values provide \cite{dressel15}, and there are those who attempt to define in which sense reality can be 
ascribed to a property of a quantum system, which leads to the ascription 
of reality in a specific sense to the weak values \cite{matzkin19,reznik20}.

Here we explain the quantum pigeonhole effect \cite{aharonov16,chen19} and the exchange of grins between quantum Cheshire cats \cite{das20,liu20} as simple interference effects. In our description, the quantum system that acts as a pointer in the weak measurement formalism is incorporated to 
an enlarged system, such that the division into system of interest and pointer is absent. And since we avoid talking about weak values at all, we can also address on equal footing cases in which the experiment does not use weak measurements to extract weak values \cite{chen19,liu20}. Under this view, the question of whether a realistic property is revealed by the weak value is suspended, because the weak values are absent from the discussion. 
The main objective is to demystify some of the conclusions of these works, such as ``quantum mechanics violates one of the fundamental principles of nature: If you put three particles in two boxes, necessarily two particles will end up in the same box'' \cite{aharonov16} or ``we report an experiment with manifoldly entangled photons that demonstrates a stronger object–property separation, whereby an object can permanently drop 
a certain property and acquire a property that it did not have from another object'' \cite{liu20}. No such conclusions appear in the descriptions presented here. Nonetheless it is important to recognize that, even without such dramatic conclusions, the observed effects are interesting in themselves. So a second objective of the present paper is to give more physical insights to them. For instance, we associate the particles anomalous behavior in the quantum pigeonhole effect with the quantum interference of force effect \cite{correa18,cenni19}.

\section{Quantum pigeonhole effect}\label{sec:pigeonhole}

Despite the fact that no mention of weak values is made in the original paper on the quantum pigeonhole effect \cite{aharonov16}, the basic concepts are there: pre- and postselection with a weak interaction among the system and a pointer, whose effect would give information about the intermediate system state. For this reason we will first describe the paradox using the weak value concept. The result is the same as in Ref. \cite{aharonov16} because the paradox is constructed to result in a null effect on the pointer. If this effect was not zero, the weak value would quantify it. 

Consider that we have three quantum particles with two possible orthogonal states $\ket{L}$ and $\ket{R}$ each, associated with their presence in two different boxes. The system is prepared in the preselected state
\begin{equation}\label{pre}
	\ket{\Psi}=\ket{+}_1\ket{+}_2\ket{+}_3, \mathrm{with}\; \ket{+}=\frac{1}{\sqrt{2}}\big[\ket{L}+\ket{R}\big],
\end{equation}
 and postselected in the state
\begin{equation}\label{pos}
	\ket{\Phi}=\ket{+i}_1\ket{+i}_2\ket{+i}_3, \mathrm{with}\; \ket{+i}=\frac{1}{\sqrt{2}}\big[\ket{L}+i\ket{R}\big].
\end{equation}
Consider that, between the pre- and the postselection above, there is a weak 
interaction between the system and a continuous-variable quantum system that acts as a pointer. The interaction is represented by a term $gAP$ in the Hamiltonian, where $A$ is an operator for the system, $P$ is an operator 
for the pointer, and $g$ is a small coupling constant. The effect of this interaction with pre- and postselection is to produce a displacement on the 
wave function of the pointer on the variable conjugate to $P$ by an amount proportional to the real part of the weak value of $A$, defined as \cite{aharonov88}
\begin{equation}
	\avg{A}_w=\frac{\bra{\Phi}A{\ket{\Psi}}}{\langle\Phi|\Psi\rangle}, \label{weakv}
\end{equation}
as long as this displacement is much smaller than the width of the wave function. The origin of the quantum paradoxes based on weak values is to attribute a physical reality to these weak values.

Consider the operator 
\begin{eqnarray}\label{pisame}
	&&\Pi^{same}_{i,j}=\Pi^{LL}_{i,j}+\Pi^{RR}_{i,j}, \; \mathrm{with}\\\nonumber
	&&\Pi^{LL}_{i,j}=\ket{L}_i\ket{L}_j\bra{L}_i\bra{L}_j\;\mathrm{and}\;\Pi^{RR}_{i,j}=\ket{R}_i\ket{R}_j\bra{R}_i\bra{R}_j,
\end{eqnarray}
where $i\neq j$. This operator is associated with the probability of finding particles $i$ and $j$ in the same box. It can be readily shown that for 
the pre- and postselected states of Eqs. (\ref{pre}) and (\ref{pos}), the weak values of $\Pi^{same}_{i,j}$ for all pairs of particles, given by Eq. (\ref{weakv}), are zero. This result led the authors of Ref. \cite{aharonov16} to conclude that ``given the above pre- and postselection, we have three particles in two boxes, yet no two particles can be found in the same box---our quantum pigeonhole principle.''

The paradoxical conclusions of the quantum pigeonhole effect were previously criticized based on logical and mathematical arguments \cite{svensson17}. Here we present physical arguments that give a clear picture of the problem, describing it from a quantum interference perspective, where no such paradox arises.

\begin{figure}
  \centering
    \includegraphics[width=8cm]{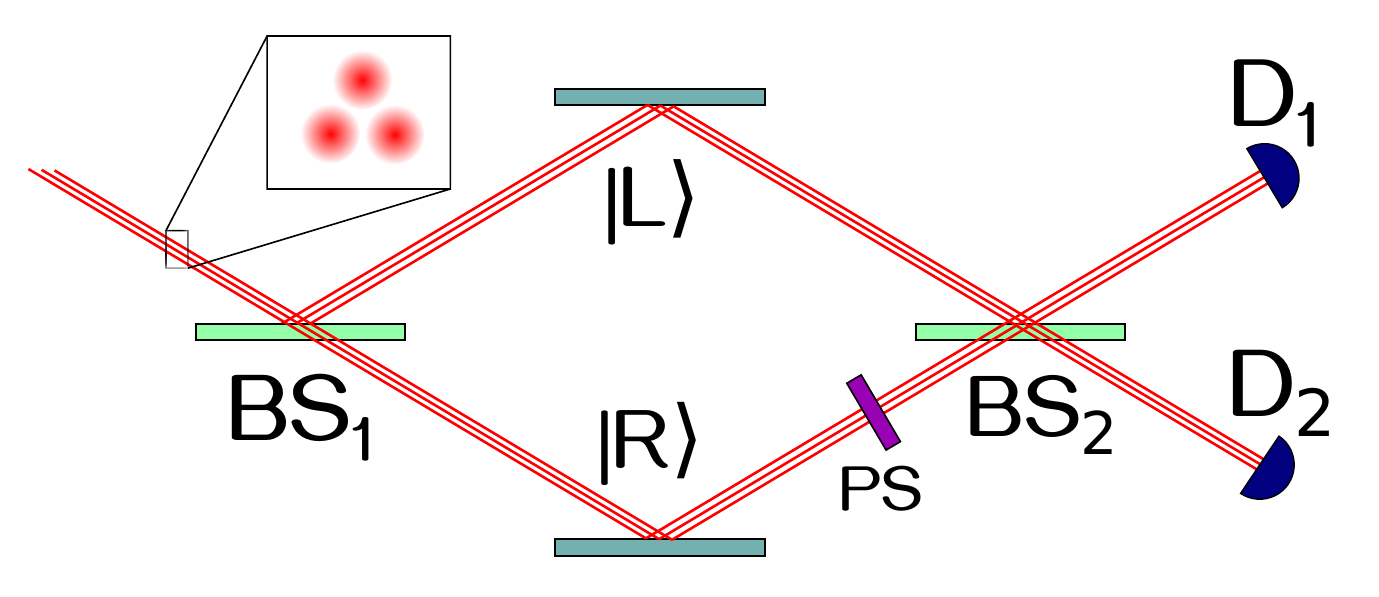}
  \caption{Scheme of Ref. \cite{aharonov16} to verify the quantum pigeonhole effect, consisting of a Mach-Zehnder interferometer with three electrons propagating parallel to each other. BS$_1$ and BS$_2$ are beam splitters, 
PS is a phase shifter and D$_1$ and D$_2$ are detectors.}\label{fig:pigeon1}
\end{figure}

In Ref. \cite{aharonov16} the scheme depicted in Fig. \ref{fig:pigeon1} is proposed to demonstrate the quantum pigeonhole effect. Three electrons are sent at the same time through a Mach-Zehnder interferometer, propagating parallel to each other. Each electron evolves to state $\ket{+}$ of Eq. (\ref{pre}) after its interaction with the beam splitter BS$_1$, where $\ket{L}$ and $\ket{R}$ label the different paths in the interferometer. The phases of the interferometer are set such that a state $\ket{+i}$ of Eq. (\ref{pos}) has constructive interference to exit in the direction of detector D$_1$, i.e., a detection by D$_1$ postselect the state $\ket{+i}$. It is assumed that in the propagation through the interferometer two electrons do not interact if they propagate in opposite paths, whereas if they propagate in the same path they suffer a weak repulsion with a final change in the relative momentum between them much smaller than the inital relative momentum uncertainty. In this sense, the relative momentum between electrons $i$ and $j$ act as a pointer for the weak measurement of the operator $\Pi^{same}_{i,j}$ from Eq. (\ref{pisame}) for the pre- and postselected states of Eqs. (\ref{pre}) and (\ref{pos}). Since the weak value of  $\Pi^{same}_{i,j}$ is zero in this situation, by selecting the situations where electrons $i$ and $j$ exit in the direction of detector D$_1$, no change in their relative momentum is noted, since this momentum change is proportional to the weak value. In Ref.  \cite{aharonov16} it is concluded that ``the beams are completely undeflected and undisturbed (up to second-order perturbations), indicating that indeed there was no interaction whatsoever between the electrons.'' As we argue in the 
following, this conclusion does not apply in an interferometric analysis.

It was recently discussed how the superposition of a positive force with a null force on a quantum particle may result in a negative momentum transfer for the particle, in an effect named quantum interference of force \cite{correa18}. This effect may result in an effective attraction between 
electrons that propagate through an interferometer, when the appropriate postselection is made \cite{cenni19}. So the interference of force may result in attraction or repulsion between electrons. In the following we discuss that the undisturbed electrons wave functions obtained in the quantum pigeonhole effect is not the result of an absence of interaction between the electrons, but the result of a particular combination of no interaction with repulsion.

Let us describe not only the electrons paths in the interferometer of Fig. \ref{fig:pigeon1}, but also their momentum wave function. Initially we have a separate state with a wave function $\Psi_1(\mb{p}_1)\Psi_2(\mb{p}_2)\Psi_3(\mb{p}_3)$, where $\mb{p}_i$ represents the momentum of electron $i$. To simplify the notation, we consider that the undisturbed electron beams always propagate in the $z$ direction, such that the coordinate system is changed with the beams reflections. With this extended representation, the system state just after BS$_1$ in Fig. \ref{fig:pigeon1} can be written as
\begin{eqnarray}
	\ket{\Psi_0}&&\propto\Big[ \ket{LLL}+\ket{LLR}+\ket{LRL}+\ket{LRR}+\ket{RLL}\\\nonumber
	                   && +\ket{RLR}+\ket{RRL}+\ket{RRR}\Big]\Psi_1(\mb{p}_1)\Psi_2(\mb{p}_2)\Psi_3(\mb{p}_3).\label{mom-init}
\end{eqnarray}
If we consider the state just before BS$_2$, due to the repulsion between electrons that propagate in the same path, the system state evolves to
\begin{eqnarray}
	\ket{\Psi_1}&&\propto\big[\ket{LLL}+\ket{RRR}\big]\Psi_1(\mb{p}_1+\bs{\delta}_{12}+\bs{\delta}_{13})\\\nonumber
	&&\;\;\;\;\;\;\;\;\;\;\;\;\;\;\;\;\;\;\;\;\times\Psi_2(\mb{p}_2-\bs{\delta}_{12}+\bs{\delta}_{23})\Psi_3(\mb{p}_3-\bs{\delta}_{13}-\bs{\delta}_{23})\\\nonumber
	                  &&+\big[\ket{LLR}+\ket{RRL}\big]\Psi_1(\mb{p}_1+\bs{\delta}_{12})\Psi_2(\mb{p}_2-\bs{\delta}_{12})\Psi_3(\mb{p}_3)\\\nonumber
										&&+\big[\ket{LRL}+\ket{RLR}\big]\Psi_1(\mb{p}_1+\bs{\delta}_{13})\Psi_2(\mb{p}_2)\Psi_3(\mb{p}_3-\bs{\delta}_{13})\\\nonumber
										&&+\big[\ket{LRR}+\ket{RLL}\big]\Psi_1(\mb{p}_1)\Psi_2(\mb{p}_2+\bs{\delta}_{23})\Psi_3(\mb{p}_3-\bs{\delta}_{23}),\label{mom-inter}
\end{eqnarray}
where $\pm\bs{\delta}_{ij}$ are small momentum displacements (perpendicular to the electrons beam propagation direction) in the wave function of electrons $i$ and $j$ when they propagate in the same path. Postselecting on the state of Eq. (\ref{pos}), which means to select the situations where all electrons exit the interferometer in the direction of D$_1$ in Fig. \ref{fig:pigeon1}, we obtain the following wave function for the electrons momenta:
\begin{eqnarray}\label{ps}\nonumber
	&\Psi_{ps}&\propto [1-i]\Psi_1(\mb{p}_1+\bs{\delta}_{12}+\bs{\delta}_{13})\Psi_2(\mb{p}_2-\bs{\delta}_{12}+\bs{\delta}_{23})\\\nonumber
	&&\;\;\;\;\;\;\;\;\;\;\;\;\;\;\;\;\;\;\;\;\;\;\;\;\;\;\;\;\;\;\;\;\;\;\;\;\;\;\;\;\times\Psi_3(\mb{p}_3-\bs{\delta}_{13}-\bs{\delta}_{23})\\\nonumber
	                && +[-1+i]\Psi_1(\mb{p}_1+\bs{\delta}_{12})\Psi_2(\mb{p}_2-\bs{\delta}_{12})\Psi_3(\mb{p}_3)\\\nonumber
									&& +[-1+i]\Psi_1(\mb{p}_1+\bs{\delta}_{13})\Psi_2(\mb{p}_2)\Psi_3(\mb{p}_3-\bs{\delta}_{13})\\
									&& +[-1+i]\Psi_1(\mb{p}_1)\Psi_2(\mb{p}_2+\bs{\delta}_{23})\Psi_3(\mb{p}_3-\bs{\delta}_{23}).									
\end{eqnarray}
Using the expansion $\Psi(\mb{r}+\bs{\delta}) \approx \Psi(\mb{r})+\bs{\delta}\cdot[\bs{\nabla}\Psi(\mb{r})]$ in Eq. (\ref{ps}) and disregarding terms with products $|\bs{\delta}_{ij}||\bs{\delta}_{kl}|$ we obtain $\Psi_{ps}\propto \Psi_1(\mb{p}_1)\Psi_2(\mb{p}_2)\Psi_3(\mb{p}_3)$, the same initial wave function. The authors of Ref. \cite{aharonov16} associated the absence of a change in the particles wave function with an absence of interaction between the electrons in the interferometer, which would demonstrate that no two electrons have propagated through the same path. But as evidenced in Eq. (\ref{ps}), this wave function is obtained by the coherent combination of different perturbed wave functions, so it is not possible to conclude that there was no interaction between 
the electrons inside the interferometer. All terms in Eq. (\ref{ps}) have at least one repulsion effect, since at least two electrons must propagate through the same interferometer arm. It is intuitively expected that the combination of repulsion with no interaction for a pair of particles should result in an effective repulsion. But as shown in Ref. \cite{cenni19}, a coherent combination of repulsion with no interaction between two quantum particles may result in an effective attraction due to an interference effect. Here we are dealing with a similar counter-intuitive situation, where the coherent combination of repulsion with no interaction results in a zero effective force. This null effective force can be verified by measuring the average relative momenta of an ensemble of pre- and postselected particles, which, according 
the obtained postselected wave function, should be unperturbed in relation to the initial system state.

\begin{figure}
  \centering
    \includegraphics[width=8cm]{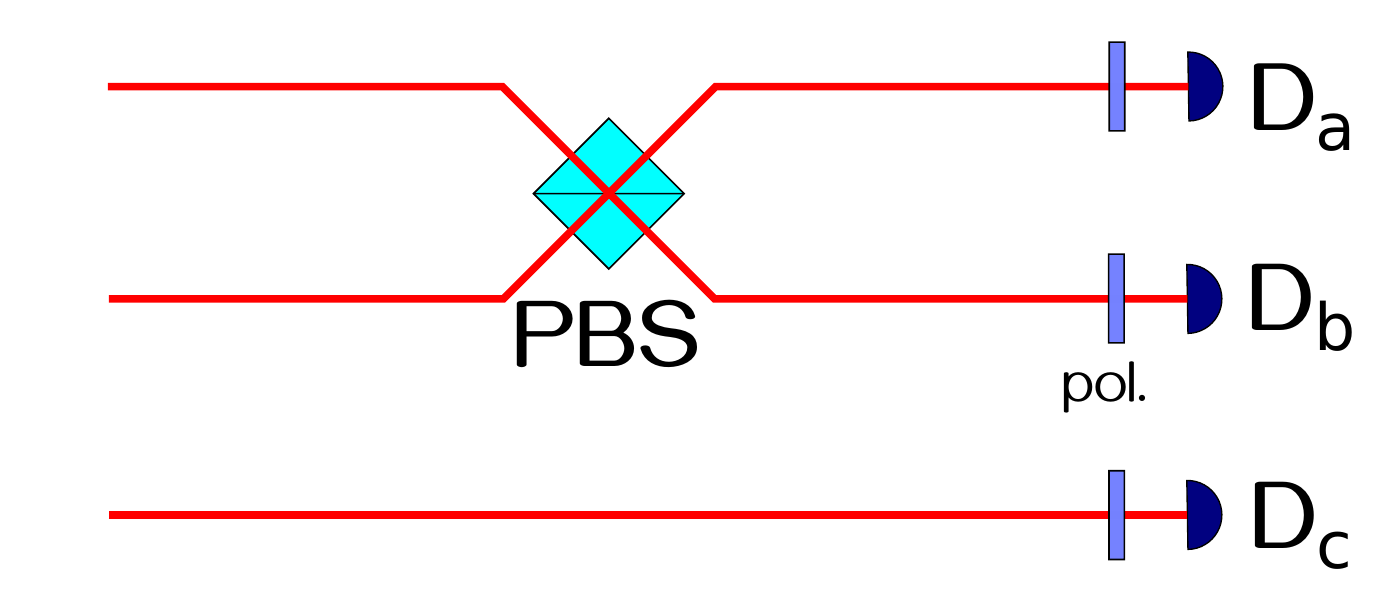}
  \caption{Experimental scheme of Ref. \cite{chen19} to show the quantum pigeonhole effect. Three photons come from the left in the indicated paths. PBS is a polarizing beam splitter, and D$_i$ are detectors with polarizers (pol.).}\label{fig:pigeon2}
\end{figure}

In an experimental implementation of the quantum pigeonhole effect \cite{chen19}, Chen \textit{et al.} use photons as the quantum ``pigeons'' and their polarization as the ``pigeonholes''. The pre- and postselected states from Eqs. (\ref{pre}) and (\ref{pos}), as well as the operator from Eq. (\ref{pisame}) that would tell if two pigeons are in the same pigeonhole, are the same, with the substitutions $\ket{L}\rightarrow\ket{H}$ and $\ket{R}\rightarrow\ket{V}$, where $\ket{H}$ is associated with a horizontal polarization for the photon and $\ket{V}$ with a vertical polarization. In the experiments they want to show that the matrix element $\bra{\Phi}\Pi^{same}_{12}\ket{\Psi}$ is zero. To do so, they perform the experiment depicted in Fig. \ref{fig:pigeon2}. Three photons (1, 2 and 3) are prepared in the polarization state $(\ket{H}+\ket{V})/\sqrt{2}$, and three detectors (D$_a$, D$_b$ and D$_c$) project the polarization state in $(\ket{H}+i\ket{V})/\sqrt{2}$. Photon 3 is sent directly to D$_c$, being detected with probability 1/2. Photons 1 and 2 are sent each to  different input ports of a polarizing beam splitter (PBS), which transmits horizontally polarized photons and reflects vertically polarized ones, whereas one output port directs the photons to D$_a$, and the other directs to D$_b$. So its use, with the selection of the situations where one photon leaves the PBS in the 
mode towards D$_a$ and the other towards D$_b$ in Fig. \ref{fig:pigeon2}, 
is associated with the action of the projection operator defined in Eq. (\ref{pisame}). They observe zero triple coincidences in this situation, demonstrating that indeed $\bra{\Phi}\Pi^{same}_{12}\ket{\Psi}=0$. By symmetry, the exchange of the roles of the photons in the experiment leads to the same result.

The experimental results of Ref. \cite{chen19} mentioned in the preceding 
paragraph can be simply understood in terms of interference. After the interaction with the PBS, the state of photons 1 and 2 is
\begin{equation}
	\frac{1}{2}\Big[\ket{H_a,H_b}+\ket{V_a,V_b}+\ket{H_a,V_a}+\ket{H_b,V_b}\Big].
\end{equation}
Detectors D$_a$ and D$_b$ project this state on
\begin{equation}
	\frac{1}{2}\Big[\ket{H_a,H_b}-\ket{V_a,V_b}+i\ket{H_a,V_b}+i\ket{V_a,H_b}\Big],
\end{equation}
resulting in a zero probability of coincidence detections. This can be simply understood in terms of Hong-Ou-Mandel interference \cite{hong87} with a PBS followed by polarization measurements. The probability amplitude of the two photons being reflected interferes destructively with the probability amplitude of them being transmitted in this situation. 

Chen \textit{et al.} also consider a more elaborate interferometer in their work, that effectively interfere the three photons \cite{chen19}. But this is used to discuss the fact that the quantum pigeonhole effect would 
be valid only in the weak interaction regime, so we do not treat these other results here.

As shown here, the theoretical proposal \cite{aharonov16} and the experimental implementation \cite{chen19} of the quantum pigeonhole effect can be understood as simple interference effects. There is no need to say that 
the pigeonhole principle is violated in quantum systems.

\section{Exchange of grins between quantum Cheshire cats} We now proceed to interpret the `exchange of grins between quantum Cheshire cats' experiment \cite{liu20}, based on the theoretical proposal \cite{das20},  from a quantum interference point of view. In these works, the Cheshire cats are photons and their grins are the photon polarizations.

\begin{figure}
  \centering
    \includegraphics[width=8cm]{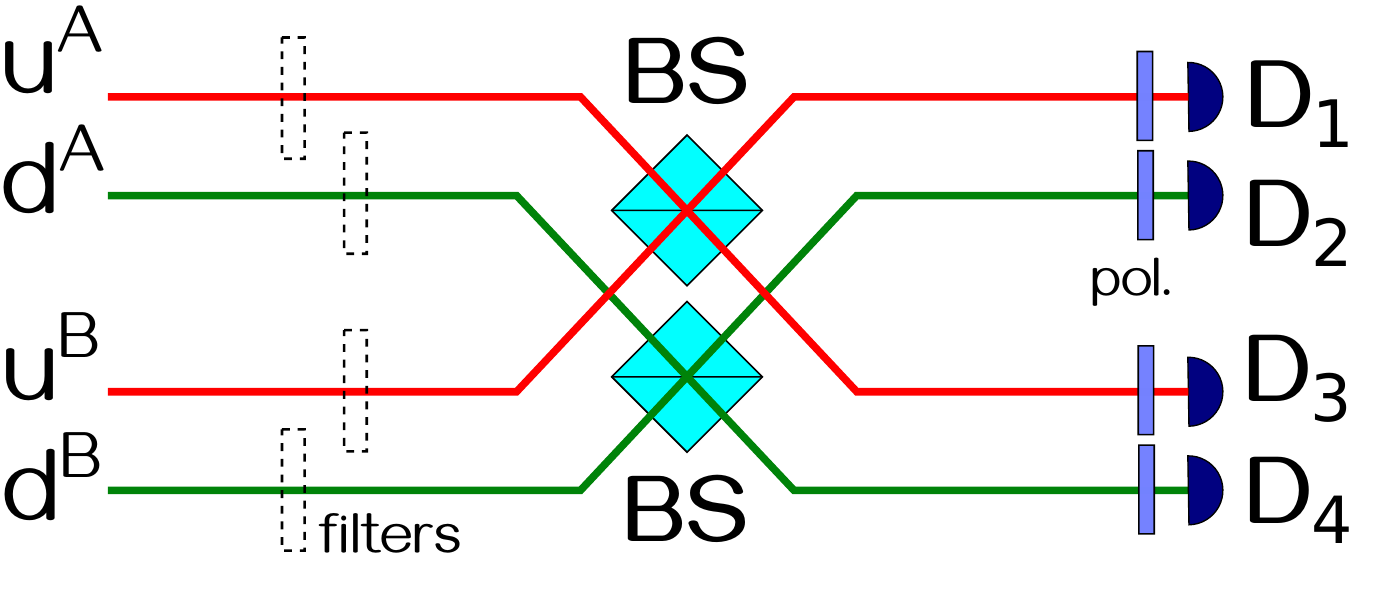}
  \caption{Scheme of the `exchange of grins between quantum Cheshire cats' experiment \cite{liu20}. Photon pairs are prepared in the initial state 
of Eq. (\ref{pre2}). Beam splitters BSs and polarizers (pol.) make a coincidence count by detectors D$_1$ and D$_2$ to postselect the state of Eq. (\ref{pos2}). Optical filters that may or may not depend on polarization can be inserted in the regions indicated by the traced rectangles to associate the weak values of Eq. (\ref{cheshire}) with the change in the coincidence counts, according to Eq. (\ref{weakprob}).}\label{fig:cheshire}
\end{figure}

A simplified experimental scheme is illustrated in Fig. \ref{fig:cheshire} (see Ref. \cite{liu20} for the details). Photon pairs of the same frequency are produced by parametric down conversion in the following entangled state
\begin{eqnarray}\nonumber
	\ket{\Psi}=\frac{1}{{2}}\Big\{&-&\Big[\ket{\uparrow^A\uparrow^B}-\ket{\downarrow^A\downarrow^B}\Big]\ket{u^Ad^B} \\
	&+&\Big[\ket{\uparrow^A\uparrow^B}+\ket{\downarrow^A\downarrow^B}\Big]\ket{d^Au^B}\Big\},\label{pre2}
\end{eqnarray}
where $\ket{u^i}$ and $\ket{d^i}$ are the path states indicated in Fig. \ref{fig:cheshire} and $\ket{\uparrow^i}$ and $\ket{\downarrow^i}$  represent horizontal and vertical polarization states respectively, for photons 
labeled $A$ and $B$.  Polarizers in the detectors project on the polarization state $(\ket{\uparrow}+\ket{\downarrow})/\sqrt{2}$. Considering that 
the path to detector D$_1$ projects on the path state $(\ket{u^A}+\ket{u^B})/\sqrt{2}$ and the path to detector D$_2$ projects on the path state $(-\ket{d^A}+\ket{d^B})/\sqrt{2}$, a double photon detection by D$_1$ and D$_2$ corresponds to a postselection of the state
\begin{eqnarray}\label{phi_2}\nonumber
	\ket{\Phi}=\frac{1}{{4}}&&\Big[\Big(\ket{\uparrow^A}+\ket{\downarrow^A}\Big)\Big(\ket{\uparrow^B}+\ket{\downarrow^B}\Big)\Big]\\
   \otimes &&\Big[\Big(\ket{u^A}+\ket{u^B}\Big)\Big(-\ket{d^A}+\ket{d^B}\Big)\Big].\label{pos2}
\end{eqnarray}
The probability of a coincidence detection by detectors D$_1$ and D$_2$ in this situation is then given by $|\langle \Phi|\Psi\rangle|^2 = 1/16$.

 The paradox arises when computing the weak values for the observables $\Pi_\mu^\nu$ and $\sigma_z^\nu\otimes\Pi_\mu^\nu$, with $\mu=\{u,d\}$ and 
$\nu=\{A,B\}$, $\Pi_u^\nu=\proj{u^\nu}$, $\Pi_d^\nu=\proj{d^\nu}$, and $\sigma_z^\nu=\proj{\uparrow^\nu}-\proj{\downarrow^\nu}$. The operator $\Pi_\mu^\nu$ has eigenvalues 1 and 0. An eigenvalue 1 (0) indicates the presence (absence) of photon $\nu$ in path $\mu^\nu$. The operator $\sigma_z^\nu\otimes\Pi_\mu^\nu$ has eigenvalues 0 and $\pm1$. An eigenvalue +1 ($-1$) indicates the presence of photon  $\nu$ in path $\mu^\nu$ with a polarization $H$ $(V)$. An eigenvalue $0$ indicates the absence of photon $\nu$ in path $\mu^\nu$. For the pre- and postselected states of Eqs. (\ref{pre2}) and (\ref{pos2}), the weak value for these operators can be computed using Eq. (\ref{weakv}), and are found to be
\begin{align}
	\avg{\Pi_u^A}_w= 0,&\ \ \ \ \avg{\sigma_z\otimes\Pi_u^A}_w=1,\notag\\
	\avg{\Pi_d^A}_w=1,&\ \ \ \ \avg{\sigma_z\otimes\Pi_d^A}_w= 0,\notag\\
	\avg{\Pi_u^B}_w=1,&\ \ \ \ \avg{\sigma_z\otimes\Pi_u^B}_w= 0,\notag\\
	\avg{\Pi_d^B}_w= 0,&\ \ \ \ \avg{\sigma_z\otimes\Pi_d^B}_w=1.\label{cheshire}
\end{align}
The authors of Ref. \cite{liu20} interpret the first two lines of the above equations as an indication that photon $A$ is in path $d$, whereas its polarization is in path $u$. The last two lines would indicate that photon 
$B$ is in path $u$, whereas its polarization is in path $d$. On top of that, the superposition of the $u$ ($d$) paths for the detection would mean that photon $B$ ($A$) is detected with photon $A$'s ($B$'s) polarization. Based on that, they conclude that ``each of the two quantum Cheshire cats 
deterministically swaps grin with its counterparts'' \cite{liu20}. But note that this conclusion is based on a realistic interpretation of the weak values, associating an objective reality with the properties of quantum particles in superposition states, which is a very controversial assumption. 

To extract the weak values in the experiment of Ref. \cite{liu20}, optical filters are positioned in specific points of the setup and their influence on the detection statistics is analyzed. These filters, modeled as a nonunitary evolution $\exp(-Ot)$ for an observable $O$, have the effect of decreasing the amplitude of certain components of the system's state, and they may or may not be polarization-sensitive. The probability for preparing a system in state $\ket{\Psi}$, the filter acting for a small time interval $t$, and performing a postselection in state $\ket{\Phi}$, is
\begin{align}
P(O,t) &= |\bra{\Phi}\exp(-Ot)\ket{\Psi}|\approx |\bra{\Phi}1-Ot\ket{\Psi}|^2 \notag\\
&\approx\left|\braket{\Phi}{\Psi}\left[1-t\frac{\bra{\Phi}O\ket{\Psi}}{\braket{\Phi}{\Psi}}\right]\right|^2\notag\\
&\approx\left|\braket{\Phi}{\Psi}\right|^2\left[1-2t\mathrm{Re}\avg{O}_w\right],
\label{weakprob}
\end{align}
where Eq. (\ref{weakv}) was used. So the real part of the weak value is proportional to the first-order derivative of $P(O,t)$ with respect to the 
parameter $t$, when terms of order $t^2$ and higher can be neglected \cite{liu20,dressel14}. 

To  experimentally obtain the weak values for $\Pi_\mu^\nu$, optical filters that reduce the light amplitude by a known amount are positioned in each path indicated in Fig. \ref{fig:cheshire} and the change in the probability of postselection is compared to Eq. (\ref{weakprob}) \cite{liu20}. Optical filters that reduce the vertical polarization component of light by a known amount are also positioned in each path indicated in Fig. \ref{fig:cheshire}. The change in the probability of postselection of Eq. (\ref{weakprob}) can then be associated with the weak value of the operator $\proj{\downarrow^\nu}\otimes\Pi_\mu^\nu$. The weak value of $\sigma_z^\nu\otimes\Pi_\mu^\nu$ is then obtained from these measurements due to 
the following relation:
\begin{equation}\label{sigma}
    \sigma_z^\nu\otimes\Pi_\mu^\nu = \Pi_\mu^\nu -2 \proj{\downarrow^\nu}\otimes\Pi_\mu^\nu,
\end{equation}
with an equivalent expression relating the weak values of these quantities.

The experimental results can be simply understood in terms of two-photon interference. Consider the measurement of the weak value of  $\Pi_u^A$. Inserting an optical filter in path $u^A$ subtracts the initial quantum state of Eq. (\ref{pre2}) by an amount
\begin{equation}
    t\Pi_u^A\ket{\Psi}=-\frac{t}{2}
    \Big[\ket{\uparrow^A\uparrow^B}-\ket{\downarrow^A\downarrow^B}\Big]\ket{u^Ad^B}.
\end{equation}
The scalar product of the above reduction with the postselected state of 
Eq. (\ref{pos2}) results in $\bra{\Phi} t\Pi_u^A\ket{\Psi}=0$. This scalar product is proportional to the numerator of the weak value of $\Pi_u^A$ given by Eq. (\ref{weakprob}) (the denominator is given by $\langle\Phi|\Psi\rangle=-1/4$), and for this reason we have $\avg{\Pi_u^A}_w= 0$, as indicated in Eq. (\ref{cheshire}). So, according to Eq. (\ref{weakprob}), the inclusion of this optical filter does not change the detection 
coincidence counts by detectors D$_1$ and D$_2$. The reason is simple: We 
are removing a portion of the initial state which is not postselected anyway. 

Consider now that we insert a polarization-dependent optical filter in path $u^A$ that subtracts the initial quantum state of Eq. (\ref{pre2}) by an amount
\begin{equation}
    t\Big(\proj{\downarrow^A}\otimes\Pi_u^A\Big)\ket{\Psi}=\frac{t}{2}
    \ket{\downarrow^A\downarrow^B}\ket{u^Ad^B}.
\end{equation}
The scalar product of the above reduction with the postselected state of 
Eq. (\ref{pos2}) results in $\bra{\Phi}  \big(t\proj{\downarrow^A}\otimes\Pi_u^A\big)\ket{\Psi}=t/8$, such that we have $\avg{\big(\proj{\downarrow^A}\otimes\Pi_u^A\big)}_w= 1/2$ and, according to Eq. (\ref{sigma}), 
$\avg{\sigma_z\otimes\Pi_u^A}_w=1$, as indicated in Eq. (\ref{cheshire}). So, according to Eq. (\ref{weakprob}), the inclusion of this polarization-dependent optical filter does change the coincidence counts by detectors D$_1$ and D$_2$. This happens because we are removing a portion of the initial state which is postselected by this double detection. 

Similar reasoning applies to all weak values indicated in Eq. (\ref{cheshire}). In this way, the experimental results from Ref. \cite{liu20} can be interpreted as simple two-photon interference. There is no need to say that the photons have exchanged their spins in the interferometer.

\section{Discussion and Conclusion}

As mentioned before, in the experiments that implemented the `quantum pigeonhole effect' \cite{chen19} and the `exchange of grins between Cheshire cats' \cite{liu20} there were no genuine weak measurement procedures, with the weak interaction of the pertinent systems degrees-of-freedom with pointers, whose shifts would reveal the weak values of the analyzed quantities. Instead, other interferometric procedures were performed to extract the weak values, that were explained here as simple interference effects. It is important to stress that, even if genuine weak measurements related to these paradoxes are performed, it would still be possible to explain the experimental results as interference effects, as we did in the case of the original proposal of the quantum pigeonhole effect \cite{aharonov16} in Sec. \ref{sec:pigeonhole}. It is always possible to interpret the experimental results of weak measurements as interference phenomena affecting the pointers' states, without attributing reality to the weak values, as done here and in previous works that demystified other paradoxes based on weak values \cite{correa15,atherton15,saldanha14,bartkiewicz15,englert17}.

To conclude, we have shown that the quantum pigeonhole effect and the exchange of grins between quantum Cheshire cats can be explained as the result of simple quantum interference. The dramatic conclusions that the pigeonhole principle is violated in quantum systems or that two particles can exchange properties without meeting each other only appear if we adopt a realistic view of the weak values. One can then choose among giving a realistic interpretation to the weak values, generating a large number of paradoxes \cite{aharonov13,denkmayr14,vaidman13,danan13,aharonov16,waegell17,chen19,reznik20,das20,liu20}, or seeing the weak values as quantities without an objective reality, avoiding such paradoxical behaviors. We prefer the second option. Our paper also gives physical insights for the origin of strange behaviors of quantum systems in these situations, for instance, associating the quantum pigeonhole effect to a quantum interference of force \cite{correa18,cenni19}.

\begin{acknowledgments}
	This work was supported by the Brazilian agencies CNPq and FAPEMIG.
\end{acknowledgments}

%\bibliography{biblio}% Produces the bibliography via BibTeX.

\begin{thebibliography}{26}%
	\makeatletter
	\providecommand \@ifxundefined [1]{%
		\@ifx{#1\undefined}
	}%
	\providecommand \@ifnum [1]{%
		\ifnum #1\expandafter \@firstoftwo
		\else \expandafter \@secondoftwo
		\fi
	}%
	\providecommand \@ifx [1]{%
		\ifx #1\expandafter \@firstoftwo
		\else \expandafter \@secondoftwo
		\fi
	}%
	\providecommand \natexlab [1]{#1}%
	\providecommand \enquote  [1]{``#1''}%
	\providecommand \bibnamefont  [1]{#1}%
	\providecommand \bibfnamefont [1]{#1}%
	\providecommand \citenamefont [1]{#1}%
	\providecommand \href@noop [0]{\@secondoftwo}%
	\providecommand \href [0]{\begingroup \@sanitize@url \@href}%
	\providecommand \@href[1]{\@@startlink{#1}\@@href}%
	\providecommand \@@href[1]{\endgroup#1\@@endlink}%
	\providecommand \@sanitize@url [0]{\catcode `\\12\catcode `\$12\catcode
		`\&12\catcode `\#12\catcode `\^12\catcode `\_12\catcode `\%12\relax}%
	\providecommand \@@startlink[1]{}%
	\providecommand \@@endlink[0]{}%
	\providecommand \url  [0]{\begingroup\@sanitize@url \@url }%
	\providecommand \@url [1]{\endgroup\@href {#1}{\urlprefix }}%
	\providecommand \urlprefix  [0]{URL }%
	\providecommand \Eprint [0]{\href }%
	\providecommand \doibase [0]{https://doi.org/}%
	\providecommand \selectlanguage [0]{\@gobble}%
	\providecommand \bibinfo  [0]{\@secondoftwo}%
	\providecommand \bibfield  [0]{\@secondoftwo}%
	\providecommand \translation [1]{[#1]}%
	\providecommand \BibitemOpen [0]{}%
	\providecommand \bibitemStop [0]{}%
	\providecommand \bibitemNoStop [0]{.\EOS\space}%
	\providecommand \EOS [0]{\spacefactor3000\relax}%
	\providecommand \BibitemShut  [1]{\csname bibitem#1\endcsname}%
	\let\auto@bib@innerbib\@empty
	%</preamble>
	\bibitem [{\citenamefont {Aharonov}\ \emph {et~al.}(1988)\citenamefont
		{Aharonov}, \citenamefont {Albert},\ and\ \citenamefont
		{Vaidman}}]{aharonov88}%
	\BibitemOpen
	\bibfield  {author} {\bibinfo {author} {\bibfnamefont {Y.}~\bibnamefont
			{Aharonov}}, \bibinfo {author} {\bibfnamefont {D.~Z.}\ \bibnamefont
			{Albert}},\ and\ \bibinfo {author} {\bibfnamefont {L.}~\bibnamefont
			{Vaidman}},\ }\bibfield  {title} {\bibinfo {title} {How the result of a
			measurement of a component of the spin of a spin-1/2 particle can turn 
out to
			be 100},\ }\href {https://doi.org/10.1103/PhysRevLett.60.1351} {\bibfield
		{journal} {\bibinfo  {journal} {Phys. Rev. Lett.}\ }\textbf {\bibinfo
			{volume} {60}},\ \bibinfo {pages} {1351} (\bibinfo {year}
		{1988})}\BibitemShut {NoStop}%
	\bibitem [{\citenamefont {Dressel}\ \emph {et~al.}(2014)\citenamefont
		{Dressel}, \citenamefont {Malik}, \citenamefont {Miatto}, \citenamefont
		{Jordan},\ and\ \citenamefont {Boyd}}]{dressel14}%
	\BibitemOpen
	\bibfield  {author} {\bibinfo {author} {\bibfnamefont {J.}~\bibnamefont
			{Dressel}}, \bibinfo {author} {\bibfnamefont {M.}~\bibnamefont {Malik}},
		\bibinfo {author} {\bibfnamefont {F.~M.}\ \bibnamefont {Miatto}}, \bibinfo
		{author} {\bibfnamefont {A.~N.}\ \bibnamefont {Jordan}},\ and\ \bibinfo
		{author} {\bibfnamefont {R.~W.}\ \bibnamefont {Boyd}},\ }\bibfield  {title}
	{\bibinfo {title} {Colloquium: Understanding quantum weak values: Basics 
and
			applications},\ }\href {https://doi.org/10.1103/RevModPhys.86.307} {\bibfield
		{journal} {\bibinfo  {journal} {Rev. Mod. Phys.}\ }\textbf {\bibinfo
			{volume} {86}},\ \bibinfo {pages} {307} (\bibinfo {year} {2014})}\BibitemShut
	{NoStop}%
	\bibitem [{\citenamefont {Aharonov}\ \emph {et~al.}(2013)\citenamefont
		{Aharonov}, \citenamefont {Popescu}, \citenamefont {Rohrlich},\ and\
		\citenamefont {Skrzypczyk}}]{aharonov13}%
	\BibitemOpen
	\bibfield  {author} {\bibinfo {author} {\bibfnamefont {Y.}~\bibnamefont
			{Aharonov}}, \bibinfo {author} {\bibfnamefont {S.}~\bibnamefont {Popescu}},
		\bibinfo {author} {\bibfnamefont {D.}~\bibnamefont {Rohrlich}},\ and\
		\bibinfo {author} {\bibfnamefont {P.}~\bibnamefont {Skrzypczyk}},\ }\bibfield
	{title} {\bibinfo {title} {Quantum {C}heshire cats},\ }\href
	{https://doi.org/10.1088/1367-2630/15/11/113015} {\bibfield  {journal}
		{\bibinfo  {journal} {New J. Phys.}\ }\textbf {\bibinfo {volume} {15}},\
		\bibinfo {pages} {113015} (\bibinfo {year} {2013})}\BibitemShut {NoStop}%
	\bibitem [{\citenamefont {Denkmayr}\ \emph {et~al.}(2014)\citenamefont
		{Denkmayr}, \citenamefont {Geppert}, \citenamefont {Sponar}, \citenamefont
		{Lemmel}, \citenamefont {Matzkin}, \citenamefont {Tollaksen},\ and\
		\citenamefont {Hasegawa}}]{denkmayr14}%
	\BibitemOpen
	\bibfield  {author} {\bibinfo {author} {\bibfnamefont {T.}~\bibnamefont
			{Denkmayr}}, \bibinfo {author} {\bibfnamefont {H.}~\bibnamefont {Geppert}},
		\bibinfo {author} {\bibfnamefont {S.}~\bibnamefont {Sponar}}, \bibinfo
		{author} {\bibfnamefont {H.}~\bibnamefont {Lemmel}}, \bibinfo {author}
		{\bibfnamefont {A.}~\bibnamefont {Matzkin}}, \bibinfo {author} {\bibfnamefont
			{J.}~\bibnamefont {Tollaksen}},\ and\ \bibinfo {author} {\bibfnamefont
			{Y.}~\bibnamefont {Hasegawa}},\ }\bibfield  {title} {\bibinfo {title}
		{Observation of a quantum {C}heshire cat in a matter-wave interferometer
			experiment},\ }\href {https://doi.org/10.1038/ncomms5492} {\bibfield
		{journal} {\bibinfo  {journal} {Nat. Commun.}\ }\textbf {\bibinfo {volume}
			{5}},\ \bibinfo {pages} {4492} (\bibinfo {year} {2014})}\BibitemShut
	{NoStop}%
	\bibitem [{\citenamefont {Vaidman}(2013)}]{vaidman13}%
	\BibitemOpen
	\bibfield  {author} {\bibinfo {author} {\bibfnamefont {L.}~\bibnamefont
			{Vaidman}},\ }\bibfield  {title} {\bibinfo {title} {Past of a quantum
			particle},\ }\href {https://doi.org/10.1103/PhysRevA.87.052104} {\bibfield
		{journal} {\bibinfo  {journal} {Phys. Rev. A}\ }\textbf {\bibinfo {volume}
			{87}},\ \bibinfo {pages} {052104} (\bibinfo {year} {2013})}\BibitemShut
	{NoStop}%
	\bibitem [{\citenamefont {Danan}\ \emph {et~al.}(2013)\citenamefont {Danan},
		\citenamefont {Farfurnik}, \citenamefont {Bar-Ad},\ and\ \citenamefont
		{Vaidman}}]{danan13}%
	\BibitemOpen
	\bibfield  {author} {\bibinfo {author} {\bibfnamefont {A.}~\bibnamefont
			{Danan}}, \bibinfo {author} {\bibfnamefont {D.}~\bibnamefont {Farfurnik}},
		\bibinfo {author} {\bibfnamefont {S.}~\bibnamefont {Bar-Ad}},\ and\ \bibinfo
		{author} {\bibfnamefont {L.}~\bibnamefont {Vaidman}},\ }\bibfield  {title}
	{\bibinfo {title} {Asking photons where they have been},\ }\href
	{https://doi.org/10.1103/PhysRevLett.111.240402} {\bibfield  {journal}
		{\bibinfo  {journal} {Phys. Rev. Lett.}\ }\textbf {\bibinfo {volume} {111}},\
		\bibinfo {pages} {240402} (\bibinfo {year} {2013})}\BibitemShut {NoStop}%
	\bibitem [{\citenamefont {Aharonov}\ \emph {et~al.}(2016)\citenamefont
		{Aharonov}, \citenamefont {Colombo}, \citenamefont {Popescu}, \citenamefont
		{Sabadini}, \citenamefont {Struppa},\ and\ \citenamefont
		{Tollaksen}}]{aharonov16}%
	\BibitemOpen
	\bibfield  {author} {\bibinfo {author} {\bibfnamefont {Y.}~\bibnamefont
			{Aharonov}}, \bibinfo {author} {\bibfnamefont {F.}~\bibnamefont {Colombo}},
		\bibinfo {author} {\bibfnamefont {S.}~\bibnamefont {Popescu}}, \bibinfo
		{author} {\bibfnamefont {I.}~\bibnamefont {Sabadini}}, \bibinfo {author}
		{\bibfnamefont {D.~C.}\ \bibnamefont {Struppa}},\ and\ \bibinfo {author}
		{\bibfnamefont {J.}~\bibnamefont {Tollaksen}},\ }\bibfield  {title} {\bibinfo
		{title} {Quantum violation of the pigeonhole principle and the nature of
			quantum correlations},\ }\href {https://doi.org/10.1073/pnas.1522411112}
	{\bibfield  {journal} {\bibinfo  {journal} {Proc. Natl. Acad. Sci. U.S.A.}\
		}\textbf {\bibinfo {volume} {113}},\ \bibinfo {pages} {532} (\bibinfo {year}
		{2016})}\BibitemShut {NoStop}%
	\bibitem [{\citenamefont {Waegell}\ \emph {et~al.}(2017)\citenamefont
		{Waegell}, \citenamefont {Denkmayr}, \citenamefont {Geppert}, \citenamefont
		{Ebner}, \citenamefont {Jenke}, \citenamefont {Hasegawa}, \citenamefont
		{Sponar}, \citenamefont {Dressel},\ and\ \citenamefont
		{Tollaksen}}]{waegell17}%
	\BibitemOpen
	\bibfield  {author} {\bibinfo {author} {\bibfnamefont {M.}~\bibnamefont
			{Waegell}}, \bibinfo {author} {\bibfnamefont {T.}~\bibnamefont {Denkmayr}},
		\bibinfo {author} {\bibfnamefont {H.}~\bibnamefont {Geppert}}, \bibinfo
		{author} {\bibfnamefont {D.}~\bibnamefont {Ebner}}, \bibinfo {author}
		{\bibfnamefont {T.}~\bibnamefont {Jenke}}, \bibinfo {author} {\bibfnamefont
			{Y.}~\bibnamefont {Hasegawa}}, \bibinfo {author} {\bibfnamefont
			{S.}~\bibnamefont {Sponar}}, \bibinfo {author} {\bibfnamefont
			{J.}~\bibnamefont {Dressel}},\ and\ \bibinfo {author} {\bibfnamefont
			{J.}~\bibnamefont {Tollaksen}},\ }\bibfield  {title} {\bibinfo {title}
		{Confined contextuality in neutron interferometry: Observing the quantum
			pigeonhole effect},\ }\href {https://doi.org/10.1103/PhysRevA.96.052131}
	{\bibfield  {journal} {\bibinfo  {journal} {Phys. Rev. A}\ }\textbf {\bibinfo
			{volume} {96}},\ \bibinfo {pages} {052131} (\bibinfo {year}
		{2017})}\BibitemShut {NoStop}%
	\bibitem [{\citenamefont {Chen}\ \emph {et~al.}(2019)\citenamefont {Chen},
		\citenamefont {Liu}, \citenamefont {Luo}, \citenamefont {Huang},
		\citenamefont {Wang}, \citenamefont {Wang}, \citenamefont {Li}, \citenamefont
		{Liu}, \citenamefont {Lu},\ and\ \citenamefont {Pan}}]{chen19}%
	\BibitemOpen
	\bibfield  {author} {\bibinfo {author} {\bibfnamefont {M.-C.}\ \bibnamefont
			{Chen}}, \bibinfo {author} {\bibfnamefont {C.}~\bibnamefont {Liu}}, \bibinfo
		{author} {\bibfnamefont {Y.-H.}\ \bibnamefont {Luo}}, \bibinfo {author}
		{\bibfnamefont {H.-L.}\ \bibnamefont {Huang}}, \bibinfo {author}
		{\bibfnamefont {B.-Y.}\ \bibnamefont {Wang}}, \bibinfo {author}
		{\bibfnamefont {X.-L.}\ \bibnamefont {Wang}}, \bibinfo {author}
		{\bibfnamefont {L.}~\bibnamefont {Li}}, \bibinfo {author} {\bibfnamefont
			{N.-L.}\ \bibnamefont {Liu}}, \bibinfo {author} {\bibfnamefont {C.-Y.}\
			\bibnamefont {Lu}},\ and\ \bibinfo {author} {\bibfnamefont {J.-W.}\
			\bibnamefont {Pan}},\ }\bibfield  {title} {\bibinfo {title} {Experimental
			demonstration of quantum pigeonhole paradox},\ }\href
	{https://doi.org/10.1073/pnas.1815462116} {\bibfield  {journal} {\bibinfo
			{journal} {Proc. Natl. Acad. Sci. U.S.A.}\ }\textbf {\bibinfo {volume}
			{116}},\ \bibinfo {pages} {1549} (\bibinfo {year} {2019})}\BibitemShut
	{NoStop}%
	\bibitem [{\citenamefont {Reznik}\ \emph {et~al.}(2020)\citenamefont {Reznik},
		\citenamefont {Bagchi}, \citenamefont {Dressel},\ and\ \citenamefont
		{Vaidman}}]{reznik20}%
	\BibitemOpen
	\bibfield  {author} {\bibinfo {author} {\bibfnamefont {G.}~\bibnamefont
			{Reznik}}, \bibinfo {author} {\bibfnamefont {S.}~\bibnamefont {Bagchi}},
		\bibinfo {author} {\bibfnamefont {J.}~\bibnamefont {Dressel}},\ and\ \bibinfo
		{author} {\bibfnamefont {L.}~\bibnamefont {Vaidman}},\ }\bibfield  {title}
	{\bibinfo {title} {Footprints of quantum pigeons},\ }\href
	{https://doi.org/10.1103/PhysRevResearch.2.023004} {\bibfield  {journal}
		{\bibinfo  {journal} {Phys. Rev. Research}\ }\textbf {\bibinfo {volume}
			{2}},\ \bibinfo {pages} {023004} (\bibinfo {year} {2020})}\BibitemShut
	{NoStop}%
	\bibitem [{\citenamefont {Das}\ and\ \citenamefont {Pati}(2020)}]{das20}%
	\BibitemOpen
	\bibfield  {author} {\bibinfo {author} {\bibfnamefont {D.}~\bibnamefont
			{Das}}\ and\ \bibinfo {author} {\bibfnamefont {A.~K.}\ \bibnamefont {Pati}},\
	}\bibfield  {title} {\bibinfo {title} {Can two quantum {C}heshire cats
			exchange grins?},\ }\href {https://doi.org/10.1088/1367-2630/ab8e5a}
	{\bibfield  {journal} {\bibinfo  {journal} {New J. Phys.}\ }\textbf {\bibinfo
			{volume} {22}},\ \bibinfo {pages} {063032} (\bibinfo {year}
		{2020})}\BibitemShut {NoStop}%
	\bibitem [{\citenamefont {Liu}\ \emph {et~al.}(2020)}]{liu20}%
	\BibitemOpen
	\bibfield  {author} {\bibinfo {author} {\bibfnamefont {Z.-H.}\ \bibnamefont
			{Liu}}\ \bibinfo {author} {\emph {et~al.}},\ }\bibfield  {title}
	{\bibinfo {title} {Experimental exchange of grins between quantum {C}heshire
			cats},\ }\href {https://doi.org/10.1038/s41467-020-16761-0} {\bibfield
		{journal} {\bibinfo  {journal} {Nat. Commun.}\ }\textbf {\bibinfo {volume}
			{11}},\ \bibinfo {pages} {3006} (\bibinfo {year} {2020})}\BibitemShut
	{NoStop}%
	\bibitem [{\citenamefont {Corr{\^{e}}a}\ \emph {et~al.}(2015)\citenamefont
		{Corr{\^{e}}a}, \citenamefont {Santos}, \citenamefont {Monken},\ and\
		\citenamefont {Saldanha}}]{correa15}%
	\BibitemOpen
	\bibfield  {author} {\bibinfo {author} {\bibfnamefont {R.}~\bibnamefont
			{Corr{\^{e}}a}}, \bibinfo {author} {\bibfnamefont {M.~F.}\ \bibnamefont
			{Santos}}, \bibinfo {author} {\bibfnamefont {C.~H.}\ \bibnamefont {Monken}},\
		and\ \bibinfo {author} {\bibfnamefont {P.~L.}\ \bibnamefont {Saldanha}},\
	}\bibfield  {title} {\bibinfo {title} {`{Q}uantum {C}heshire cat' as simple
			quantum interference},\ }\href
	{https://doi.org/10.1088/1367-2630/17/5/053042} {\bibfield  {journal}
		{\bibinfo  {journal} {New J. Phys.}\ }\textbf {\bibinfo {volume} {17}},\
		\bibinfo {pages} {053042} (\bibinfo {year} {2015})}\BibitemShut {NoStop}%
	\bibitem [{\citenamefont {Atherton}\ \emph {et~al.}(2015)\citenamefont
		{Atherton}, \citenamefont {Ranjit}, \citenamefont {Geraci},\ and\
		\citenamefont {Weinstein}}]{atherton15}%
	\BibitemOpen
	\bibfield  {author} {\bibinfo {author} {\bibfnamefont {D.~P.}\ \bibnamefont
			{Atherton}}, \bibinfo {author} {\bibfnamefont {G.}~\bibnamefont {Ranjit}},
		\bibinfo {author} {\bibfnamefont {A.~A.}\ \bibnamefont {Geraci}},\ and\
		\bibinfo {author} {\bibfnamefont {J.~D.}\ \bibnamefont {Weinstein}},\
	}\bibfield  {title} {\bibinfo {title} {Observation of a classical {C}heshire
			cat in an optical interferometer},\ }\href
	{https://doi.org/10.1364/OL.40.000879} {\bibfield  {journal} {\bibinfo
			{journal} {Opt. Lett.}\ }\textbf {\bibinfo {volume} {40}},\ \bibinfo {pages}
		{879} (\bibinfo {year} {2015})}\BibitemShut {NoStop}%
	\bibitem [{\citenamefont {Saldanha}(2014)}]{saldanha14}%
	\BibitemOpen
	\bibfield  {author} {\bibinfo {author} {\bibfnamefont {P.~L.}\ \bibnamefont
			{Saldanha}},\ }\bibfield  {title} {\bibinfo {title} {Interpreting a nested
			mach-zehnder interferometer with classical optics},\ }\href
	{https://doi.org/10.1103/PhysRevA.89.033825} {\bibfield  {journal} {\bibinfo
			{journal} {Phys. Rev. A}\ }\textbf {\bibinfo {volume} {89}},\ \bibinfo
		{pages} {033825} (\bibinfo {year} {2014})}\BibitemShut {NoStop}%
	\bibitem [{\citenamefont {Bartkiewicz}\ \emph {et~al.}(2015)\citenamefont
		{Bartkiewicz}, \citenamefont {\ifmmode~\check{C}\else \v{C}\fi{}ernoch},
		\citenamefont {Jav\ifmmode~\mathring{u}\else \r{u}\fi{}rek}, \citenamefont
		{Lemr}, \citenamefont {Soubusta},\ and\ \citenamefont
		{Svozil\'{\i}k}}]{bartkiewicz15}%
	\BibitemOpen
	\bibfield  {author} {\bibinfo {author} {\bibfnamefont {K.}~\bibnamefont
			{Bartkiewicz}}, \bibinfo {author} {\bibfnamefont {A.}~\bibnamefont
			{\ifmmode~\check{C}\else \v{C}\fi{}ernoch}}, \bibinfo {author} {\bibfnamefont
			{D.}~\bibnamefont {Jav\ifmmode~\mathring{u}\else \r{u}\fi{}rek}}, \bibinfo
		{author} {\bibfnamefont {K.}~\bibnamefont {Lemr}}, \bibinfo {author}
		{\bibfnamefont {J.}~\bibnamefont {Soubusta}},\ and\ \bibinfo {author}
		{\bibfnamefont {J.~c.~v.}\ \bibnamefont {Svozil\'{\i}k}},\ }\bibfield
	{title} {\bibinfo {title} {One-state vector formalism for the evolution of a
			quantum state through nested mach-zehnder interferometers},\ }\href
	{https://doi.org/10.1103/PhysRevA.91.012103} {\bibfield  {journal} {\bibinfo
			{journal} {Phys. Rev. A}\ }\textbf {\bibinfo {volume} {91}},\ \bibinfo
		{pages} {012103} (\bibinfo {year} {2015})}\BibitemShut {NoStop}%
	\bibitem [{\citenamefont {Englert}\ \emph {et~al.}(2017)\citenamefont
		{Englert}, \citenamefont {Horia}, \citenamefont {Dai}, \citenamefont {Len},\
		and\ \citenamefont {Ng}}]{englert17}%
	\BibitemOpen
	\bibfield  {author} {\bibinfo {author} {\bibfnamefont {B.-G.}\ \bibnamefont
			{Englert}}, \bibinfo {author} {\bibfnamefont {K.}~\bibnamefont {Horia}},
		\bibinfo {author} {\bibfnamefont {J.}~\bibnamefont {Dai}}, \bibinfo {author}
		{\bibfnamefont {Y.~L.}\ \bibnamefont {Len}},\ and\ \bibinfo {author}
		{\bibfnamefont {H.~K.}\ \bibnamefont {Ng}},\ }\bibfield  {title} {\bibinfo
		{title} {Past of a quantum particle revisited},\ }\href
	{https://doi.org/10.1103/PhysRevA.96.022126} {\bibfield  {journal} {\bibinfo
			{journal} {Phys. Rev. A}\ }\textbf {\bibinfo {volume} {96}},\ \bibinfo
		{pages} {022126} (\bibinfo {year} {2017})}\BibitemShut {NoStop}%
	\bibitem [{\citenamefont {Leggett}(1989)}]{leggett89}%
	\BibitemOpen
	\bibfield  {author} {\bibinfo {author} {\bibfnamefont {A.~J.}\ \bibnamefont
			{Leggett}},\ }\bibfield  {title} {\bibinfo {title} {Comment on ``how the
			result of a measurement of a component of the spin of a spin-(1/2 particle
			can turn out to be 100''},\ }\href
	{https://doi.org/10.1103/PhysRevLett.62.2325} {\bibfield  {journal} {\bibinfo
			{journal} {Phys. Rev. Lett.}\ }\textbf {\bibinfo {volume} {62}},\ \bibinfo
		{pages} {2325} (\bibinfo {year} {1989})}\BibitemShut {NoStop}%
	\bibitem [{\citenamefont {Sokolovski}(2015)}]{sokolovski15}%
	\BibitemOpen
	\bibfield  {author} {\bibinfo {author} {\bibfnamefont {D.}~\bibnamefont
			{Sokolovski}},\ }\bibfield  {title} {\bibinfo {title} {The meaning of
			“anomalous weak values” in quantum and classical theories},\ }\href
	{https://doi.org/https://doi.org/10.1016/j.physleta.2015.02.018} {\bibfield
		{journal} {\bibinfo  {journal} {Phys. Lett. A}\ }\textbf {\bibinfo {volume}
			{379}},\ \bibinfo {pages} {1097} (\bibinfo {year} {2015})}\BibitemShut
	{NoStop}%
	\bibitem [{\citenamefont {Sokolovski}\ and\ \citenamefont
		{Akhmatskaya}(2018)}]{sokolovski18}%
	\BibitemOpen
	\bibfield  {author} {\bibinfo {author} {\bibfnamefont {D.}~\bibnamefont
			{Sokolovski}}\ and\ \bibinfo {author} {\bibfnamefont {E.}~\bibnamefont
			{Akhmatskaya}},\ }\bibfield  {title} {\bibinfo {title} {An even simpler
			understanding of quantum weak values},\ }\href
	{https://doi.org/10.1016/j.aop.2017.11.030} {\bibfield  {journal} {\bibinfo
			{journal} {Ann. Phys. (N.Y.)}\ }\textbf {\bibinfo {volume} {388}},\ \bibinfo {pages}
		{382} (\bibinfo {year} {2018})}\BibitemShut {NoStop}%
	\bibitem [{\citenamefont {Dressel}(2015)}]{dressel15}%
	\BibitemOpen
	\bibfield  {author} {\bibinfo {author} {\bibfnamefont {J.}~\bibnamefont
			{Dressel}},\ }\bibfield  {title} {\bibinfo {title} {Weak values as
			interference phenomena},\ }\href {https://doi.org/10.1103/PhysRevA.91.032116}
	{\bibfield  {journal} {\bibinfo  {journal} {Phys. Rev. A}\ }\textbf {\bibinfo
			{volume} {91}},\ \bibinfo {pages} {032116} (\bibinfo {year}
		{2015})}\BibitemShut {NoStop}%
	\bibitem [{\citenamefont {Matzkin}(2019)}]{matzkin19}%
	\BibitemOpen
	\bibfield  {author} {\bibinfo {author} {\bibfnamefont {A.}~\bibnamefont
			{Matzkin}},\ }\bibfield  {title} {\bibinfo {title} {Weak values and quantum
			properties},\ }\href {https://doi.org/10.1007/s10701-019-00245-3} {\bibfield
		{journal} {\bibinfo  {journal} {Found. Phys.}\ }\textbf {\bibinfo {volume}
			{49}},\ \bibinfo {pages} {298} (\bibinfo {year} {2019})}\BibitemShut
	{NoStop}%
	\bibitem [{\citenamefont {Corr{\^{e}}a}\ \emph {et~al.}(2018)\citenamefont
		{Corr{\^{e}}a}, \citenamefont {Cenni},\ and\ \citenamefont
		{Saldanha}}]{correa18}%
	\BibitemOpen
	\bibfield  {author} {\bibinfo {author} {\bibfnamefont {R.}~\bibnamefont
			{Corr{\^{e}}a}}, \bibinfo {author} {\bibfnamefont {M.~F.~B.}\ \bibnamefont
			{Cenni}},\ and\ \bibinfo {author} {\bibfnamefont {P.~L.}\ \bibnamefont
			{Saldanha}},\ }\bibfield  {title} {\bibinfo {title} {Quantum interference of
			force},\ }\href {https://doi.org/10.22331/q-2018-12-14-112} {\bibfield
		{journal} {\bibinfo  {journal} {Quantum}\ }\textbf {\bibinfo {volume} {2}},\
		\bibinfo {pages} {112} (\bibinfo {year} {2018})}\BibitemShut {NoStop}%
	\bibitem [{\citenamefont {Cenni}\ \emph {et~al.}(2019)\citenamefont {Cenni},
		\citenamefont {Corr\^ea},\ and\ \citenamefont {Saldanha}}]{cenni19}%
	\BibitemOpen
	\bibfield  {author} {\bibinfo {author} {\bibfnamefont {M.~F.~B.}\
			\bibnamefont {Cenni}}, \bibinfo {author} {\bibfnamefont {R.}~\bibnamefont
			{Corr\^ea}},\ and\ \bibinfo {author} {\bibfnamefont {P.~L.}\ \bibnamefont
			{Saldanha}},\ }\bibfield  {title} {\bibinfo {title} {Effective electrostatic
			attraction between electrons due to quantum interference},\ }\href
	{https://doi.org/10.1103/PhysRevA.100.022101} {\bibfield  {journal} {\bibinfo
			{journal} {Phys. Rev. A}\ }\textbf {\bibinfo {volume} {100}},\ \bibinfo
		{pages} {022101} (\bibinfo {year} {2019})}\BibitemShut {NoStop}%
	\bibitem [{\citenamefont {Svensson}(2017)}]{svensson17}%
	\BibitemOpen
	\bibfield  {author} {\bibinfo {author} {\bibfnamefont {B.~E.~Y.}\
			\bibnamefont {Svensson}},\ }\bibfield  {title} {\bibinfo {title} {Quantum
			weak values and logic: An uneasy couple},\ }\href
	{https://doi.org/10.1007/s10701-017-0068-5} {\bibfield  {journal} {\bibinfo
			{journal} {Found. Phys.}\ }\textbf {\bibinfo {volume} {47}},\ \bibinfo
		{pages} {430} (\bibinfo {year} {2017})}\BibitemShut {NoStop}%
	\bibitem [{\citenamefont {Hong}\ \emph {et~al.}(1987)\citenamefont {Hong},
		\citenamefont {Ou},\ and\ \citenamefont {Mandel}}]{hong87}%
	\BibitemOpen
	\bibfield  {author} {\bibinfo {author} {\bibfnamefont {C.~K.}\ \bibnamefont
			{Hong}}, \bibinfo {author} {\bibfnamefont {Z.~Y.}\ \bibnamefont {Ou}},\ and\
		\bibinfo {author} {\bibfnamefont {L.}~\bibnamefont {Mandel}},\ }\bibfield
	{title} {\bibinfo {title} {Measurement of subpicosecond time intervals
			between two photons by interference},\ }\href
	{https://doi.org/10.1103/PhysRevLett.59.2044} {\bibfield  {journal} {\bibinfo
			{journal} {Phys. Rev. Lett.}\ }\textbf {\bibinfo {volume} {59}},\ \bibinfo
		{pages} {2044} (\bibinfo {year} {1987})}\BibitemShut {NoStop}%
\end{thebibliography}


\end{document}